\documentclass[prl,twocolumn]{revtex4}
\usepackage{amsmath}
\usepackage{graphicx}
\usepackage{amsfonts}
\usepackage{amsmath}
\usepackage{amsfonts}
\usepackage{amssymb}
\begin{document}
\title{Fractional Quantum Hall States in Fast Rotating Bose Gases}
\author{A. \surname{Lakhoua},  T. Masson, J.C. \surname{Wallet}}

\affiliation{Laboratoire de Physique Th\'{e}orique, B\^{a}timent 210, F-91405 Orsay CEDEX, France}
\begin{abstract} 
We use a Chern Simons Landau-Ginzburg (CSLG) framework related to hierarchies of composite bosons to describe 2D harmonically trapped fast rotating Bose gases in Fractional Quantum Hall Effect (FQHE) states. The predicted values for $\nu$ (ratio of particle to vortex numbers) are $\nu$$=$${{p}\over{q}}$ ($p$, $q$ are any integers) with even product $pq$, including numerically favored values previously found and predicting a richer set of values. We show that those values can be understood from a bosonic analog of the law of the corresponding states relevant to the electronic FQHE. A tentative global phase diagram for the bosonic system for $\nu$$<$$1$ is also proposed.
\vskip 0,4 true cm

\pacs:{PACS numbers: 03.75.Kk, 73.43.-f}

\end{abstract}
\maketitle


Since the experimental realization of Bose-Einstein condensation (BEC) of atomic gases \cite{BEC}, an intense activity has been focused on ultracold atomic Bose gases in rotating harmonic traps. BEC confined in two dimensions have been created \cite{GORLITZ}. The rotation of a BEC produces vortices \cite{BIBIVORT}. When the rotation frequency (${{\Omega}\over{2\pi}})$ increases, the BEC is destroyed and for ${{\Omega}\over{2\pi}}$ high enough a Fractional FQHE \cite{PRANGE} state is expected to occur. In particular, when ${{\Omega}\over{2\pi}}$ is tuned to the frequency of the harmonic confining potential in the radial plane (${{\Omega_T}\over{2\pi}}$), FQHE states have been predicted to become possible ground states for the system \cite{PREDICFQHE}, triggering studies on FQHE for bosons \cite{FQHE1}-\cite{CASA}. Indeed, the Hamiltonian in the rotating frame \cite{DALF} is \cite{FOOT1} $H$$=$$\sum_{A=1}^N{{1}\over{2m}}[{\bf{p}}_A$$-$${\bf{A}}({\bf{x}}_A)]^2$$+$${{m}\over{2}}(\Omega_T^2$$-$$\Omega^2)|{\bf{x}}_A|^2$$+$$V$ (${\bf{x}}_A$$\equiv$${\bf{x}}_A(t)$);  $A_i({\bf{x}}_A)$$=$$m\Omega\epsilon_{ij}x_A^j$ yields the Coriolis force,  $V$$\simeq$$g_2\sum_{A<B}\delta({\bf{x}}_{A}$$-$${\bf{x}}_B)$ is the potential felt by the bosons in the plane \cite{FOOT2}. We take $g_2$$>$$0$ (repulsive case \cite{NOUS}), $g_2$$\simeq$$\gamma l^2$ \cite{FOOT2} with  $\gamma$$=$${\sqrt{32\pi}}\Omega({{a_s}\over{l_z}})$, $l^2$$=$${{{1}\over{2m\Omega}}}$,  $l_z^2$$=$${{{1}\over{m\Omega_z}}}$ ($a_s$, $l$, (resp.$l_z$) are the 3D $s$-wave scattering length and the localization length in the plan (resp. for axial confinement with trapping frequency $\Omega_z$)). We assume $\Omega$$=$$\Omega_T$ so that the system is equivalent to that of 2D (bosonic) particles in a magnetic field with cyclotron frequency $\Omega_c$$=$$2\Omega$, filling factor $\nu$$=$${{N}\over{N_V}}$, $2\pi N_V$$=$$2m\Omega{\cal{A}}$ \cite{FOOT3} (${\cal{A}}$ is the total area of the system) and is expected to be in FQHE states at specific values for $\nu$ \cite{PREDICFQHE}-\cite{CASA}.\par
The low energy behavior of the Quantum Hall Fluid (QHF) states can be described by a CSLG theory \cite{REVIEW} where the statistics of the quasiparticles is controlled by the coupling of a statistical CS gauge potential (hereafter called $a_\mu$) to matter \cite {REVIEW}, \cite{LERDA}. Two equivalent CSLG descriptions of the {\it{electronic}} FQHE appeared, depending whether the initial Fermi statistics of the system is turned into a Bose one through statistical transmutation \cite{LERDA} as in \cite{ZHANG} or instead kept unchanged \cite{FRADKIN}. In this note, we describe 2D harmonically trapped fast rotating Bose gases in FQHE states by a CSLG theory related to hierarchies of composite bosons. We find that FQHE states should occur at $\nu$$=$${{p}\over{q}}$  for {\it{any}} integers $p$, $q$ with even product $pq$, including the numerically favored values presented in \cite{JOLIC}, \cite{PAREDES},\cite{CASA} and predicting a richer set of values. We show that those values can be understood from a bosonic analog of the discrete symmetry transformations underlying the law of the corresponding states \cite{ZHANG} relevant to the electronic FQHE. A tentative global phase diagram for the bosonic system together with selection rules for the expected transitions are also proposed.\par
The relevant CSLG action obtained by adapting \cite{REVIEW}, \cite{ZHANG} to $H$ given above is
$$S=S_{CS}(a;\eta)+\int_xi\phi^\dag D_0\phi-{{1}\over{2m^*}}|D_i\phi|^2-U(\phi)\eqno(1),$$
$$S_{CS}(a;\eta)=\int_x{{\eta}\over{4}}
\epsilon_{\mu\nu\rho}a^\nu f^{\nu\rho};\ D_\mu=\partial_\mu-i{\tilde{A_\mu}} \eqno(2a,b).$$
$S$$\equiv$$S_{CS}(a;\eta)$$+$$S_\phi(\phi;{\tilde{A}})$, ${\tilde{A}}_\mu$$=$$a_\mu$$+$$A_\mu$.  $\phi$$=$${\sqrt{\rho}}$$e^{i\lambda(x)}$$e^{i\chi(x)}$ is the composite boson field with (effective) mass $m^*$ ($\lambda(x)$ is single valued, $\chi(x)$$=$$\sum_Vw_V\arctan({{x_2-X_{V2}}\over{x_1-X_{V1}}})$ is the possible vortex contribution \cite{VORTPHASE}; $w_V$ is the winding number of the vortex $V$ with center coordinates $X_{Vi}(t)$), $\eta$ is the CS parameter,  $f_\mu$$\equiv$${{1}\over{2}}\epsilon_{\mu\nu\rho}f^{\nu\rho}$, $f_{\mu\nu}$$=$$\partial_\mu a_\nu$$-$$\partial_\nu a_\mu$, $f_0$ ($f_i$) is the statistical magnetic (electric) field. $U(\phi)$ should be $U(\phi)$$=$$g_2(\phi^\dag\phi)^2$$-$$\mu|\phi|^2$$+$$...$($\mu$ is a chemical potential, the ellipses denote possible higher order terms). The equations of motion for $a_\mu$, $f_\mu$$=$$-$$(1/\eta)J_\mu$ ($J_0$$=$$\rho$, $J_i$$=$${{i}\over{2m}}(\phi^\dag D_i\phi$$-$$(D_i\phi)^\dag\phi)$) ensure the formation of particle-statistical magnetic flux composite particles. This reflects the flux attachment mechanism triggered by the CS statistical interaction \cite{LERDA}.  The statistics of the quasiparticles wave function is altered by the Aharonov-Bohm phase  $\exp(i\int_{\cal{C}}{\bf{a}}.d{\bf{x}})$$=$$\exp(i/\eta)$ (${\cal{C}}$ is some closed curve) induced when one quasiparticle moves around another one \cite{LERDA}. The flux attachment leaves the statistics unchanged (e.g bosons are turned into bosons) when $\eta$$=$${{1}\over{2\pi2k}}$, $k$$\in$$\mathbb{Z}$, $k$$>$$0$, that we assume from now on  \cite{TRANSMUT} while (1) corresponds to QHF with filling factor $\nu$$=$$2\pi\eta$ (see \cite{FOOT4} and \cite{FOOT5}).\par
The low energy behavior for the system in a FQHE state is encoded into the response functions (we consider the linear response) for the corresponding QHF, obtained from the effective action \cite{REVIEW}, \cite{VORTPHASE} $\Gamma_\eta(A)$$\equiv$$-$$i\ln Z_\eta(A)$ where
$$Z_\eta(A)=\int[{\cal{D}}a][{\cal{D}}\phi][{\cal{D}}\phi^\dag]e^{iS_{CS}(a;\eta)}e^{iS_\phi(\phi;{\tilde{A}})} \eqno(3).$$
In (3), we set $Z({\tilde{A}})$$\equiv$$\exp(i\Gamma({\tilde{A}}))$$=$$\int[{\cal{D}}\phi][{\cal{D}}\phi^\dag]
e^{iS_\phi(\phi;{\tilde{A}})}$ where $\Gamma({\tilde{A}})$ is the effective action for the composite bosons $\phi$ feeling an external field ${\tilde{A}}_\mu$. Hierarchies of observable values for $\nu$ can be obtained from (1)-(3) in a way similar to the one followed for the electronic QHE \cite{REVIEW}, \cite{ZHANG}, i.e from duality transformations relating QHF's with different filling factors. These transformations express the fact that the attachment of an even number of statistical flux quanta ($\eta$$=$${{1}\over{2\pi2k}}$) turns one QHF onto another one having similar physics (the flux attachment transformation) or stem from the duality transformation \cite{FISHER} leading to a description of the system based on vortices (the (quasi)particle-vortex duality)\cite{REVIEW}. We now present a compact derivation inspirated from \cite{ZHANG} of the transformations relevant to the present bosonic system.\par
Integrating over $\phi$, $\phi^\dag$ in $\exp(i\Gamma({\tilde{A}}))$, one infers that for any external ${\tilde{A_\mu}}$ the most general low energy form for $\Gamma({\tilde{A}})$ consistent with gauge and Galilean invariance and involving the most relevant (lowest dimensional) terms must be $\Gamma({\tilde{A}})$$=$$\int_{x,y}{{1}\over{2}}{\tilde{F}}_{0i}^xQ_1^{x,y}{\tilde{F}}_{0i}^y$$-$$
{{1}\over{2}}{\tilde{F}}_{12}^xQ_2^{x,y}{\tilde{F}}_{12}^y$ $-$${{1}\over{2}}\epsilon^{\mu\nu\rho}{\tilde{A}}_\mu^xQ_3^{x,y}\partial_\nu {\tilde{A}}_\rho^y$. The $Q_I$'s ($Q_I^{x,y}$$=$$Q_I(x$$-$$y$), $I$$=$$1,2,3$) are the response functions for the composite boson fluid feeling an external ${\tilde{A}}_\mu$ and ${\tilde{F}}_{\mu\nu}$$=$$\partial_\mu {\tilde{A}}_\nu$$-$$\partial_\nu {\tilde{A}}_\mu$. This gives in momentum space 
$$\Gamma({\tilde{A}})={{1}\over{2}}\int dp {\tilde{A}}_\mu(-p){{\chi}}^{\mu\nu}(p){\tilde{A}}_\nu(p) \eqno(4a),$$
$$\chi_{00}={\bf{p}}^2Q_1(p); \chi_{\substack{0i\\ (i0)}}=\omega p_iQ_1(p){\substack{+\\(-)}}i\epsilon_{ij}p^jQ_3(p) \eqno(4b;c),$$
$$\chi_{ij}=Q_1(p)\omega^2\delta_{ij}-(\delta_{ij}{\bf{p}}^2-p_ip_j)Q_2(p)+
i\epsilon_{ij}\omega Q_3(p) \eqno(4d)$$
where $p$$=$$(\omega,{\bf{p}}$$=$$(p_1,p_2)$). Plugging (4) into (3), fixing the gauge freedom \cite{GAUGEFIX}, the integration over $a_\mu$ realizes the attachment of an even number of flux quanta to the initial composite bosons, turning the corresponding parent fluid described by the $Q_I$'s into a descendant fluid. This gives $\Gamma_\eta(A)$$=$${{1}\over{2}}\int dp A_\mu(-p){\cal{K}}^{\mu\nu}(p)A_\nu(p)$ where ${\cal{K}}^{\mu\nu}$ is deduced from (4) by replacing the $Q_I$'s by 
$$\Pi_I=\eta^2{{Q_i}\over{Q_1Q_2{\bf{p}}^2-Q_1^2\omega^2+(Q_3+\eta)^2}},\ I=1,2 \eqno(5a),$$
$$\Pi_3=\eta-\eta^2{{Q_3+\eta}\over{Q_1Q_2{\bf{p}}^2-Q_1^2\omega^2+(Q_3+\eta)^2}} \eqno(5b).$$
The dual description of the system makes use of the vortices \cite{FISHER}. One applies a Hubbard-Stratonovitch (HS) transformation on the kinetic term in (1) (set $\phi$$=$${\sqrt{\rho}}e^{i\lambda}v$, $v$$=$$e^{i\chi}$). The action becomes $S_H$$=$$S_{CS}(a;\eta)$$+$$\int_xi\rho v^*\partial_0v$$-$$\rho(\partial_0\lambda$$-$$a_0)$$-$${{1}\over{2m}}(\partial_i{\sqrt{\rho}})^2$$-$
${{m}\over{2\rho}}H_i^2$$+$$H_i(\partial_i\lambda$$-$${\tilde{A}}_i$$-$$iv^*\partial_iv)$ ($H_i$ are the HS fields). Set $\rho$$\equiv$$H_0$. The integration over $\lambda$ \cite{FISHER} yields the constraint $\partial_\mu H^\mu$$=$$0$ in the partition function. It is solved by introducing a new gauge potential $b_\mu$ such that $H_\mu$$=$$\epsilon_{\mu\nu\rho}\partial^\nu b^\rho$$\equiv$$H_\mu(b)$. This yields
$$S_H^\prime=S_{CS}(a;\eta)+\int_xb_\mu {\cal{J}}^\mu+a_\mu H^\mu(b)+A_iH^i(b)$$
$$-{{m}\over{2H_0(b)}}H_i^2(b)
-{{1}\over{2m}}(\partial_i{\sqrt{H_0(b)}})^2-U(H_0(b)) \eqno(6).$$
${\cal{J}}_\mu$$=$$i\epsilon_{\mu\nu\rho}\partial^\nu v^*\partial^\rho v$ is the vortex current. The integration over $a_\mu$ leads to (${\tilde{b}}_\mu$$=$$b_\mu$$-$$\eta A_\mu$, ${\tilde{H}}_\mu$$=$$\epsilon_{\mu\nu\rho}\partial^\nu {\tilde{b}}^\rho$$=$$H_\mu({\tilde{b}})$)
$$S_D=\int_x{{1}\over{2\eta}}\epsilon_{\mu\nu\rho}b^\mu\partial^\nu b^\rho-{{\eta}\over{2}}\epsilon_{\mu\nu\rho}A^\mu\partial^\nu A^\rho+{\tilde{b}}_\mu {\cal{J}}^\mu
-{{m}\over{2{\tilde{H}}_0}}{\tilde{H}}_i^2$$
$$-{{1}\over{8m{\tilde{H}}_0}}(\partial_i{\tilde{H}}_0)^2-U({\tilde{H}}_0)
\equiv S_{CS}(b;{{1}\over{\eta}})+S_D^\prime \eqno(7),$$
where $b_\mu$ plays the role of the new statistical field while the flux attachment to the vortices is achieved thanks to the occurrence of the minimal coupling of $b_\mu$ to ${\cal{J}}_\mu$ (see 3rd term in (7)).
The partition function now reads $$Z_\eta(A)=\int[{\cal{D}}b]e^{iS_{CS}(b;{{1}\over{\eta}}))}e^{-iS_{CS}(A;\eta)}e^{i\Gamma({\tilde{b}})} \eqno(8),$$ 
with $\exp(i\Gamma({\tilde{b}}))$$=$$\int[{\cal{D}}v][{\cal{D}}v^*]\exp(S_D^\prime|_{\eta=0})$. By integrating over the vortex part $v,v^*$, gauge invariance and Galilean invariance dictate the most general low energy expression for $\Gamma({\tilde{b}})$: $\Gamma({\tilde{b}})$$=$$\int_{x,y}\big({{1}\over{2}}{\tilde{B}}_{0i}(x)V_1^{x,y}{\tilde{B}}_{0i}(y)$$-$$
{{1}\over{2}}{\tilde{B}}_{12}(x)V_2^{x,y}{\tilde{B}}_{12}(y)$\\$
-$${{1}\over{2}}\epsilon_{\mu\nu\rho}{\tilde{b}}^\mu(x)V_3^{x,y}\partial^\nu {\tilde{b}}^\rho(y)\big)$ ($V_I^{x,y}$$=$$V_I(x$$-$$y)$, $I$$=$$1,2,3$, ${\tilde{B}}_{\mu\nu}$$=$$\partial_\mu{\tilde{b}}_\nu$$-$$\partial_\nu{\tilde{b}}_\mu$). The $V_I$'s are the response functions for the vortices feeling an effective external field ${\tilde{b}}$; $V_3$ is related to the effective filling factor for the vortices $\nu_V$ via $\lim_{\omega\to0;{\bf{p}}\to0}({{V_3}\over{2\pi}})$$=$$\nu_V$. In momentum space, $\Gamma({\tilde{b}})$$=$${{1}\over{2}}\int dp {\tilde{b}}^\mu(-p){{\chi}}_{\mu\nu}^V(p){\tilde{b}}^\nu(p)$, with ${{\chi}}_{\mu\nu}^V(p)$ deduced from (4) through $Q_I$$\to$$V_I$. Combining $\Gamma({\tilde{b}})$  with (8), integrating over $b_\mu$ \cite{GAUGEFIX}, one obtains the effective action $\Gamma_\eta(A)$$=$${{1}\over{2}}\int dp A_\mu(-p){\cal{K}}^{\mu\nu}(p)A_\nu(p)$ in terms of the $V_I$'s
$$\Pi_I={{V_I}\over{V_1V_2{\bf{p}}^2-V_1^2\omega^2+(V_3-{{1}\over{\eta}})^2}},\ I=1,2 \eqno(9a),$$
$$\Pi_3={{{{1}\over{\eta}}-V_3}\over{V_1V_2{\bf{p}}^2-V_1^2\omega^2+(V_3-{{1}\over{\eta}})^2}} \eqno(9b).$$
In the low energy limit, (5b) and (9b) give rise to the flux attachment and particle-vortex duality transformations leading to hierarchies of observable values for $\nu$. These transformations generate a discrete symmetry group. It can be viewed as the bosonic analog of the discrete symmetry group related to the "law of the corresponding states" (see 2nd of \cite{ZHANG}) occurring in the CSLG description of the electronic QHE \cite{REVIEW}. Both discrete groups are conjugate to each other. We set $\lim_{\omega\to0;{\bf{p}}\to0}X(\omega,{\bf{p}})$$\equiv$${{X^0}\over{2\pi}}$, for $X$$=$$\Pi_3,Q_3$. Recall that if the $\Pi_I$'s (resp. $Q_I$'s) describe a descendant (resp. parent) QHF with filling factor $\nu_f$ (resp. $\nu$), one has $\lim_{\omega\to0;{\bf{p}}\to0}\Pi_3(\omega,{\bf{p}})$$=$${{\nu_f}\over{2\pi}}$, 
(resp. $\lim_{\omega\to0;{\bf{p}}\to0}Q_3(\omega,{\bf{p}})$$=$${{\nu}\over{2\pi}}$) \cite{FOOT7}. The $X^0$'s will be identified in a while with the corresponding filling factors but are kept unrestricted to positive values for the moment. Now, if the $Q_I$'s and $V_I$'s correspond to QHE states for the system (incompressible QHF), (5) and (9) reduce in the low energy limit respectively to $\Pi_3^0$$=$${{Q_3^0}\over{2kQ_3^0+1}}$ (i) and $\Pi_3^0$$=$$(2k$$-$${{1}\over{2\pi}}\lim_{\omega\to0;{\bf{p}}\to0}V_3)^{-1}$ (ii) (using $\eta$$=$${{1}\over{2\pi2k}}$).  This requires $\lim_{\omega\to0;{\bf{p}}\to0}({{V_3}\over{2\pi}})$$=$$-$${{1}\over{Q_3^0}}$ (iii) so that $\{(i)$, $(ii)\}$ are equivalent to $\{(i),$ $(iii)\}$. Furthermore, (i) is obtained by iterating $k$ time the transformation $Q_3^0$$\to$${{Q_3^0}\over{2Q_3^0+1}}$ so that the whole set of transformations giving the hierarchies of possible values for $Q_3^0$ are obtained by forming all the products of powers of the two independent generators $\Sigma$ and ${\cal{S}}$ defined by
$$\Sigma(Q_3^0)={{Q_3^0}\over{2Q_3^0+1}};\ {\cal{S}}(Q_3^0)=-{{1}\over{Q_3^0}}\eqno(10a;b).$$
(10a) represents the flux attachment transformation when $Q_3^0$ is identified with $\nu$\cite{REVIEW}, \cite{ZHANG}. (10b) is related to the (quasi)particle-vortex duality transformation. When acting on the complex plane, $\Sigma$ and ${\cal{S}}$ generate an infinite discrete group, a subgroup of the full modular group PSL(2,$\mathbb{Z}$) \cite{RANKIN}, known in mathematics as $\Gamma_\theta(2)$ (see \cite{RANKIN}). Any transformation $G$$\in$$\Gamma_\theta(2)$ can be written as
$$G(z)={{az+b}\over{cz+d}},\ ad-bc=1,\  (a,b,c,d)\in\mathbb{Z} \eqno(11),$$
with $(a,d)$ odd (resp. even) and $(b,c)$ even (resp. odd), for any $z$$\in$$\mathbb{C}$. $\Gamma_\theta(2)$ can be viewed as the bosonic analog of the discrete symmetry group underlying the "law of the corresponding states" \cite{ZHANG} of the electronic QHE \cite{REVIEW}. This latter is another subgroup of PSL$(2,\mathbb{Z})$, known as $\Gamma_0(2)$ \cite{RANKIN}, generated by $\Sigma$ (10a) and the Landau level shift operator ${\cal{T}}(z)$$=$$z$$+$$1$ (see \cite{ZHANG}, \cite{REVIEW}). Note that for any $\tau$$\in$$\Gamma_\theta(2)$, one can find some $\sigma$$\in$$\Gamma_0(2)$ such that
$\tau$$=$$\Lambda\sigma\Lambda^{-1}$ (iv) with $\Lambda$$=$${\cal{S}}{\cal{T}}{\cal{S}}$ \cite{RANKIN}: $\Gamma_\theta(2)$, relevant for fast rotating harmonically trapped Bose gases in FQHE regime, is conjugate of the discrete group underlying the law of the corresponding states \cite{ZHANG} of the electronic QHE, suggesting that properties of the bosonic system could be simply deduced from their electronic QHE counterparts by conjugation (iv), as discussed in a while. \par
Now, given a QHF of reference with filling factor $\nu_0$, values for the filling factors of descendant QHF are obtained by applying on $\nu_0$ successive transformations of $\Gamma_\theta(2)$ (with action restricted to the real axis), keeping in mind that only positive (or zero) values for these filling factors are physically allowed. Using (11), the hierarchies of values can be generically parametrized as
$$\{\nu\}_{{{even}\over{odd}}}={{2\alpha\nu_0+2\beta+\nu_0}\over{2\gamma\nu_0+2\delta+1}}, 2(\beta\gamma-\alpha\delta)=\alpha+\delta \eqno(12a),$$
$$\{\nu\}_{{{odd}\over{even}}}={{2\alpha\nu_0+2\beta+1}\over{2\gamma\nu_0+2\delta-\nu_0}}, 2(\alpha\delta-\beta\gamma)=\beta+\gamma \eqno(12b),$$
$(\alpha,\beta,\gamma,\delta)$$\in$$\mathbb{Z}$, (12a) (resp. (12b)) corresponds in (11) to $(b,c)$ even, $(a,d)$ odd (resp. $(b,c)$ odd, $(a,d)$ even) and the physically allowed values must be positive or zero.\par
Let us discuss the physical implications of our analysis. For small $\nu$, says $\nu$$\lesssim$$1$, FQHE states are expected to occur in the system \cite{DOUC}, as supported by \cite{JOLIC}, \cite{CASA}, \cite{PAREDES} (in particular by numerical works based on exact diagonalization). These later provide convincing numerical indications that (gapped) incompressible states with $\nu$$=$${{1}\over{m}}$, $m$ even integer, should be visible in experiments together with (gapped) incompressible states with $\nu$$=$${{p}\over{p+1}}$, for $p$ integer (with $\nu$$=$${{1}\over{2}}$ exactly realized for hard core bosons). For $1$$\lesssim$$\nu$$\lesssim$$6$, gapped states with $\nu$$=$${{3}\over{2}},{{5}\over{2}}, {{7}\over{2}}, {{9}\over{2}}, {{4}\over{3}}, {{5}\over{4}}$ are also numerically observed. It appears that these numerically favored values can be all reproduced within the present CSLG description. Namely, assuming $\nu_0$$=$${{1}\over{2}}$ (see \cite{FOOTCOMP}), these are recovered from (12a,b) or equivalently from successive $\Gamma_\theta(2)$ transformations acting on $\nu_0$$=$${{1}\over{2}}$. Indeed, $\Sigma^k(\nu_0)$$=$${{\nu_0}\over{2k\nu_0+1}}$so that $\Sigma^k({{1}\over{2}})$$=$${{1}\over{2(k+1)}}$$=$${{1}\over{m}}$, $m$ even ($k$$\ge$$0$) while values such as $\nu$$=$${{p}\over{p+1}}$, $p$ integer, are recovered from $(\Sigma{\cal{S}})^p(\nu_0)$ ($\Sigma$ and ${\cal{S}}$ defined in (10)). Similarly, the $\Gamma_\theta(2)$ transformations leading to $\nu$$=$${{3}\over{2}},{{5}\over{2}},{{7}\over{2}},{{9}\over{2}}$ are respectively $G^{3/2}(\nu_0)$$=$${{5\nu_0-4}\over{4\nu_0-3}}$, $G^{5/2}(\nu_0)$$=$$\nu_0+2$,
$G^{7/2}(\nu_0)$$=$${{13\nu_0-10}\over{4\nu_0-3}}$, $G^{9/2}(\nu_0)$$=$$\nu_0+4$.\par
The above observed agreement suggests that the transformations encoded in $\Gamma_\theta(2)$ may have captured global properties of the bosonic system in a way similar to what happens within electronic QHE whose properties connected in particular with the global phase diagram seem to be well encoded in the fermionic counterpart of $\Gamma_\theta(2)$ \cite{ZHANG}, \cite{REVIEW}. Assuming that this picture is correct, then properties of the fast rotating harmonically trapped Bose gases in FQHE regime can be simply deduced from their electronic QHE counterpart by applying the conjugation (iv). This however should be presumably valid at least sufficiently far away from the region where the vortex lattice has just melted so that we further assume $\nu$$<$$1$. In this way, we find, using (11), (12), (iv) that QHF states should be observed in fast rotating trapped Bose gases at filling factor $\nu$$=$${{p}\over{q}}$, where $p$, $q$ are {\it{any}} integers with {\it{even}} product $pq$ (the conjugate values of the observed odd denominator filling factors in electronic QHE). Furthermore, applying the conjugation (iv) to the global phase diagram of \cite{ZHANG} and to the corresponding selection rules derived from the the fermionic counterpart of $\Gamma_\theta(2)$ yields respectively the tentative phase diagram depicted on the figure 1 and the associated selection rules. Note that the state $\nu$$=$${{1}\over{2}}$ of the electronic QHE corresponds to the state $\nu$$=$$1$ in the bosonic system. We also find that transitions relating two QHF states indexed by $\nu_1$$=$${{p_1}\over{q_1}}$ and $\nu_2$$=$${{p_2}\over{q_2}}$ are allowed provided $|p_1q_2$$-$$p_2q_1|$$=$$1$ with (${{p_1}\over{q_1}}$, ${{p_2}\over{q_2}}$) of the form (${{even}\over{odd}}$,${{odd}\over{even}}$) or 
(${{odd}\over{even}}$, ${{even}\over{odd}}$). 
\begin{figure}
\includegraphics[width=\linewidth]{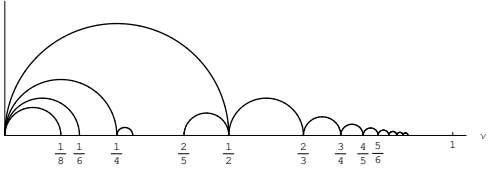}
\caption{Topology of the global phase diagram for harmonically trapped Bose gases in FQHE regime obtained from the bosonic analog of the law of the corresponding states. The vertical (resp. horizontal) axis is related to dissipation (resp. filling factor restricted here to be $\nu$$<$$1$ ). }
\label{lafigure}
\end{figure}


\begin{thebibliography}{30} 
\bibitem{BEC}: M.H. Anderson et al., Science 269 (1995), 198; K.B. Davis et al., Phys. Rev. Lett. 75 1995) 3969; W. Ketterle, Phys. Today 52 (2000) 30.
\bibitem{GORLITZ}: A. G\" orlitz et al., Phys. Rev. Lett. 87 (2001) 130402.
\bibitem{BIBIVORT}: M.R. Matthews et al., Phys. Rev. Lett. 83 (1999) 2498; K.W. Madison, F. Chevy, W. Wohlleben, J. Dalibard, Phys. Rev. Lett. 84 (2000) 806; J.R. Abo-Shaeer, C. Raman, J.M. Vogels, W. Ketterle, Science 292 (2001) 476.
\bibitem{PRANGE}: {\it{The Quantum Hall Effect}}, R.E. Prange, S.M. Girvin Eds. (Springer-Verlag, Berlin, 1990).
\bibitem{PREDICFQHE}: N.K. Wilkin, J.M.F. Gunn, R.A. Smith, Phys. Rev. Lett. 80 (1998) 2265; N.R. Cooper, N.K. Wilkin, Phys. Rev. B60 (1999) R16279; N.K. Wilkin, J.M.F. Gunn, Phys. Rev. Lett. 84 (2000) 6.
\bibitem{FQHE1}: A.D. Jackson, G.M. Kavoulakis, B.M. Helson, S.M. Reimann, Phys. Rev. Lett. 86 (2001) 945; B. Paredes, P. Fedichev, J.I. Cirac, P. Zoller, {\it{ibid.}} 87 (2001) 010402; T.-L. Ho, {\it{ibid.}} 87 (2001) 060403; U.R. Fischer, G. Baym, {\it{ibid.}} 90 (2003) 140402.
\bibitem{FQHE2}: M. Manninen, S. Viefers, M. Koskinen, S. M. Reimann, Phys. Rev. B64 (2001) 245322; J. Sinova, C. B. Hanna, A.H. MacDonald, Phys. Rev. Lett. 89 (2002) 030403; V. Sweikhard et al., {\it{ibid.}} 92 (2004) 040404. 
\bibitem{PAREDES}: B. Paredes, J.I. Cirac, P. Zoller, Phys. Rev. A66 (2002) 033609; see also B. Paredes, J.J. Garcia-Ripoll, P. Zoller, J.I. Cirac, J. Phys. IV France 116 (2004) 135.
\bibitem{JOLIC}: N. Regnault, T. Jolicoeur, Phys. Rev. Lett. 91 (2003) 30402, Phys. Rev. B 69 (2004) 235309; N.R. Cooper, N.K. Wilkin, J.M.F. Gunn, Phys. Rev. Lett. 87  (2001) 120405. 
\bibitem{CASA}: See also M.A. Cazalilla, Phys. Rev. A67 (2003) 63613.
\bibitem{DALF}: F. Dalfovo, S. Giorgini, L.P. Pitaevskii and S. Stringari, Rev. Mod. Phys. 71 (1999) 463; A. J. Leggett, {\it{ibid.}} 73 (2001) 307.)
\bibitem{FOOT1}: $h$$=$$c$$=$$1$; $x$$=$$(t,{\bf{x}})$, $\epsilon_{12}$$=$$1$; spacetime
metric is $g_{\mu\nu}$$=$diag$(+,-,-)$, $\mu,\nu,...$$=$$0,1,2$, $\epsilon_{012}$$=$$1$, 
$\partial_i$$\equiv$${{\partial}\over{\partial x^i}}$, 
$\partial_0$$\equiv$$\partial_t$, $\int_x$$\equiv$$\int
dtd{\bf{x}}$. Einstein summation is understood. 
\bibitem{NOUS}: For $g_2$$>$$0$ see U.R. Fischer, Phys. Rev. Lett. 93 (2004) 160403; A. Lakhoua, M. Lassaut, T. Masson and J.C. Wallet, Phys. Rev. A73 (2006) 023614.
\bibitem{FOOT2}: Tightening the axial confinement applied to a 3D system yields 2D system for which $g_2$ receives a contribution from the axial motion of the atoms; it can be neglected if ${{a_s}\over{l_z}}$ is small enough, that we assume, giving $g_2$$\simeq$$\gamma l^2$ ; see e.g D.S. Petrov, D.M. Gangardt and G. V. Shlyapnikov, J. Phys. IV 116 (2004) 5.
\bibitem{FOOT3}: $N_V$, the total number of vortices, plays the role of the total degeneracy of a Landau level. 
\bibitem{REVIEW}: For a review, see e.g A. Karlhede, S.A. Kivelson and S. Sondhi, in {\it{Correlated Electron Systems}}, V.J. Emery Ed., World Scientific, Singapore (1993).
\bibitem{LERDA}: A. Lerda, Lecture Notes in Physics, Vol.14 (1992); see also A. Khare, World Scientific, Singapore, 1997.
\bibitem{ZHANG}: S.-C. Zhang, T. H. Hanson and S. Kivelson, Phys. Rev. Lett. 62 (1989) 82; S. Kivelson, D.-H. Lee and S.-C. Zhang, Phys. Rev. B46 (1992) 2223.
\bibitem{FRADKIN}: A. Lopez and E. Fradkin, Phys. Rev. B44 (1991) 5246, Phys. Rev. Lett. 69 (1992) 2126.
\bibitem{VORTPHASE}: N. Nagaosa, {\it{Quantum Field Theory in Condensed Matter Physics}}, (Springer Verlag, 1999); H. Kleinert, {\it{Gauge Fields in Condensed Matter}}, (World Scientific, 1989).
\bibitem{TRANSMUT}: Flux attachement transmutes bosons into fermions (or vice versa) when $\eta$$=$${{1}\over{2\pi(2k+1)}}$ as in \cite{ZHANG}.
\bibitem{FOOT4}: The equations of motion for (1) (up to higher order terms in $U$) support the uniform density (mean field) solution $|\phi|^2$$=$${{\mu}\over{2g_2}}$, ${\bf{a}}$$+$${\bf{A}}$$=$$0$, $a_0$$=$$0$ which combined with $f_0$$=$$-$${{1}\over{\eta}}\rho$ yields $\nu$$=$$2\pi\eta$ and $\nu$$=$${{1}\over{2k}}$ for $\eta$$=$${{1}\over{2\pi2k}}$. Around this configuration, (1) yields a dispersion relation with a gap so that the fluid is incompressible \cite{REVIEW}.
\bibitem{FOOT5}: In the Gaussian approximation, (1) reduces to a set of harmonic oscillators with ground state (at low energy) equal to the Laughlin wave function $\Psi_{\nu}$ \cite{REVIEW}, \cite{VORTPHASE}.
\bibitem{MACDO}: S.M. Girvin, A.H. MacDonald, Phys. Rev. Lett. 58 (1987) 1252.
\bibitem{GAUGEFIX}: It is convenient to use the Coulomb gauge.
\bibitem{FISHER}: M.P.A. Fisher, D-H. Lee, Phys. Rev. B39 (1989) 2756; D-H. Lee, M.P.A. Fisher, Phys. Rev. Lett. 63 (1989) 903; D.H. Lee, S.-C. Zhang, Phys. Rev. Lett. 66 (1991) 1220.
\bibitem{FOOT7}: The 1st relation is obtained e.g from the definition of $\nu$ combined to $J_0$ ($J_\mu$$=$${{\delta\Gamma_\eta(A)}\over{\delta A_\mu}}$) and $A_\mu$$=$$(0$$; A_i$$=$$m\Omega\epsilon_{ij}x^j)$. A similar derivation holds for the 2nd relation.
\bibitem{RANKIN}: R.A. Rankin, {\it{Modular forms and functions}}, Cambridge University Press (1977).
\bibitem{DOUC}: For a review see e.g B. Dou\c cot, J. Phys. IV France 116 (2004) 171 and references therein.
\bibitem{FOOTCOMP}: For $\nu_0$$=$${{1}\over{2}}$, $\Phi$ in (1) should be interpreted as related to the composite bosons for the QHF with filling factor ${{1}\over{2}}$.

\end{thebibliography}
\end{document}